\documentclass[prc,aps,a4paper,nofootinbib,showkeys,showpacs,twocolumn,floatfix]{revtex4}
\usepackage{graphicx}
\usepackage{hyperref}
\usepackage{amssymb,amsfonts}
\usepackage{amsmath}
\usepackage{epstopdf}
\usepackage{natbib}
\usepackage{color}

\newcommand{\maxi}{{\text{max}}}
\newcommand{\Skyrme}{{\text{Skyrme}}}

\newcommand{\HF}{{\text{HF}}}
\newcommand{\ph}{{\text{ph}}}
\newcommand{\pp}{{\text{pp}}}
\newcommand{\ex}{{\text{ex}}}
\newcommand{\lowk}{{\text{low-}k}}
\newcommand{\mean}[1]{\langle #1\rangle}
\newcommand{\sumk}[1]{\langle\!\langle{#1}\rangle\!\rangle}

\newcommand{\calH}{\mathcal{H}}
\newcommand{\calR}{\mathcal{R}}
\newcommand{\calW}{\mathcal{W}}

\newcommand{\Rz}{\calR^{(0)}}
\newcommand{\Ru}{\calR^{(1)}}
\newcommand{\Hz}{\calH^{(0)}}
\newcommand{\Hu}{\calH^{(1)}}

\newcommand{\dRz}{\dot{\calR}^{(0)}}
\newcommand{\dRu}{\dot{\calR}^{(1)}}
\newcommand{\tRz}{\widetilde{\calR}^{(0)}}
\newcommand{\tRu}{\widetilde{\calR}^{(1)}}
\newcommand{\tHz}{\widetilde{\calH}^{(0)}}
\newcommand{\tHu}{\widetilde{\calH}^{(1)}}

\newcommand{\Fig}[1]{Fig.~\ref{#1}}
\newcommand{\Ref}[1]{Ref.~\cite{#1}}
\newcommand{\Refs}[1]{Refs.~\cite{#1}}
\newcommand{\Eq}[1]{Eq.~(\ref{#1})}
\newcommand{\Eqs}[1]{Eqs.~(\ref{#1})}
\newcommand{\vek}[1]{\mathbf{#1}}
\newcommand{\vk}{\vek{k}}
\newcommand{\vkp}{\vek{k}'}
\newcommand{\vp}{\vek{p}}
\newcommand{\vq}{\vek{q}}
\newcommand{\vqp}{\vek{q}'}
\DeclareMathOperator{\Imag}{Im}
\DeclareMathOperator{\Real}{Re}

\begin{document}
\title{Collective Modes in a Superfluid Neutron Gas within the
  Quasiparticle Random-Phase Approximation}
\author{No\"el Martin}
\email{noelmartin@ipno.in2p3.fr}
\author{Michael Urban}
\email{urban@ipno.in2p3.fr}

\affiliation{Institut de Physique Nucl\'eaire, CNRS/IN2P3 and
  Universit\'e Paris-Sud 11, F-91406 Orsay Cedex, France}
\begin{abstract}
 We study collective excitations in a superfluid neutron gas at zero
 temperature within the quasiparticle random phase approximation. The
 particle-hole residual interaction is obtained from a Skyrme
 functional, while a separable interaction is used in the pairing
 channel which gives a BCS gap that is very similar to the one
 obtained with a realistic nucleon-nucleon interaction. In accordance
 with the Goldstone theorem, we find an ungapped collective mode
 (analogous to the Bogoliubov-Anderson mode). At low momentum, its
 dispersion relation is approximately linear and its slope coincides
 with the hydrodynamic speed of sound calculated with the Skyrme
 equation of state. The response functions are compared with those
 obtained within the Landau approximation. We also compute the
 contribution of the collective mode to the specific heat of the
 neutron gas.
\end{abstract}

\maketitle
\section{Introduction}
In the inner crust of neutron stars, very neutron-rich nuclei are
immersed in a gas of unbound neutrons \cite{Negele}. A few minutes
after the formation of the neutron star, it has already cooled down
below the superfluid transition temperature $T_c$ of the neutron gas,
i.e., the neutrons form Cooper pairs. This strongly suppresses the
neutron contribution to the specific heat at low temperatures $T< T_c$
\cite{Fortin-Margueron}. However, it was pointed out that the
contribution of collective modes to the specific heat can be very
important \cite{KhanSandulescu2005}. In particular the contribution of
acoustic phonons with long wavelengths is dominant over that of gapped
neutron quasiparticles at low temperatures \cite{DiGallo}. 

In the present paper we will restrict ourselves to a simplified
system, namely a uniform neutron gas. Collective excitations in
uniform neutron and nuclear matter have been extensively studied
within the random-phase approximation (RPA)
\cite{Garcia,Pastore2012}. Collectivity may, e.g., strongly affect the
neutrino mean-free path \cite{Margueron-Neutrino}. In ordinary RPA,
however, pairing between the neutrons is not included. The extension
of RPA which accounts for pairing is called the quasiparticle RPA
(QRPA). Calculations with pairing in neutron matter \cite{Keller} and
in $\beta$-stable neutron-proton-electron ($npe$) matter
\cite{BaldoDucoin2011} (as it exists in the neutron-star core) have
been performed within the Landau approximation. But it is known from
RPA calculations \cite{Garcia} that results obtained within the Landau
approximation can differ substantially from those obtained with the
full residual particle-hole (ph) interaction derived from the Skyrme
functional. One of the goals of the present paper is to perform a full
QRPA calculation where the same Skyrme interaction that is used for
the description of the ground state is also used as residual
interaction among the quasiparticles.

In the pairing channel, we use a separable interaction that is a good
approximation to a low-momentum effective interaction ($V_{\lowk}$)
obtained by renormalization-group techniques from a realistic
nucleon-nucleon force \cite{Bogner2007}. For the sake of consistency,
we use the same interaction in the gap equation and in the
particle-particle (pp) channel of the QRPA. This guarantees that the
QRPA correctly describes the Bogoliubov-Anderson sound
\cite{Bogoliubov,Anderson}, which is a density wave with linear
dispersion relation at low momenta. This mode is actually a Goldstone
mode \cite{Goldstone} related to the broken $U(1)$ symmetry in the
superfluid phase. Note that similar calculations have been performed
in other fields of physics, e.g., ultracold atoms \cite{Combescot}.

We find that the speed of sound coincides with the hydrodynamic one
that can be calculated from the Skyrme equation of state (EOS). We
calculate the contribution of the sound mode to the specific heat and
find that it is much bigger than that of thermally excited neutron
quasiparticles.

The important role of the Goldstone mode in the neutron star crust was
already studied in numerous recent papers,
e.g. \cite{Aguilera2009,Chamel2011,DiGallo,Cirigliano2011,
  Chamel2013,Kobyakov2013}. However, in these studies the Goldstone
mode was generally treated in the long-wavelength limit and its
coupling to the two-quasiparticle continuum was neglected. This
coupling, which has already been found to be important, e.g., in the
case of ultracold atoms \cite{Combescot,Forbes2013}, is automatically
included in the QRPA.

The paper is organized as follows. In Sec.~\ref{sec:Formalism}, we
briefly explain the formalism we use to describe the ground state and
the collective modes of neutron matter. In Sec.~\ref{sec:Results}, we
discuss numerical results, and Sec.~\ref{sec:Conclusion} is devoted to
the summary and conclusions. Some technical details are given in the
Appendix.

Throughout the article, we use units with $\hbar = c = k_B = 1$
($\hbar = $ reduced Planck constant, $c = $ speed of light, $k_B = $
Boltzmann constant).
\section{Formalism}
\label{sec:Formalism}
\subsection{Skyrme energy density functional}
Let us start by briefly summarizing the description of neutron matter using the
Skyrme energy-density functional (EDF). The Skyrme functionals \cite{Brink} have
been fitted to a large variety of nuclear data. In addition, in order to be more
predictive for neutron-rich nuclei, they have also been fitted to the equation
of state of neutron matter \cite{Chabanat,ChabanatSLy4}. In the case of pure
neutron matter, the energy density can be written as
   \begin{multline}
      \mathcal{E}_\Skyrme = \frac{1}{2m} \tau + \frac{s_0}{4} \rho^2 
         + \frac{s_3}{24} \rho^{\alpha+2} 
         + \frac{s_1+ 3s_2}{8} (\rho \tau - j^2) \\ 
         + 3 \, \frac{s_1-s_2}{16} (\nabla \rho)^2 \, ,
      \label{eq:skyrmeedf}
   \end{multline}
with parameters $s_{0, \ldots , 3}$ and $\alpha$ which are defined in
Appendix~\ref{apx:skyrmeParameters}. In \Eq{eq:skyrmeedf}, $\rho$
denotes the number density of neutrons ($\rho = \rho_n$), $\tau$ is
the kinetic energy density (multiplied by $2m$, where $m$ is the
neutron mass), and $\vek{j}$ is the current. In terms of the density
matrix
\begin{equation}
\rho_{\vk,\vkp} =
  \mean{a^\dag_{\vkp\uparrow}a^{\phantom{\dag}}_{\vk\uparrow}}\,,
\label{eq:rhoOperator}
\end{equation}
where $a$ and $a^\dagger$ denote, respectively, neutron annihilation
and creation operators, these quantities are defined as
\begin{subequations}
  \begin{align}
    \rho(\vek{r}) &= 2\sum_{\vk,\vkp} \rho_{\vk,\vkp}
      e^{i(\vk-\vkp)\cdot\vek{r}}\,,\\
    \tau(\vek{r}) &= 2\sum_{\vk,\vkp} \vk\cdot\vkp \rho_{\vk,\vkp}
      e^{i(\vk-\vkp)\cdot\vek{r}}\,,\\
    \vek{j}(\vek{r}) &= \sum_{\vk,\vkp} (\vk+\vkp)\rho_{\vk,\vkp}
      e^{i(\vk-\vkp)\cdot\vek{r}}\,.
  \end{align}
\end{subequations}
Here we have assumed that the density matrices for both spin
projections ($\uparrow,\downarrow$) are equal. The term proportional
to $j^2$ in \Eq{eq:skyrmeedf} is necessary to ensure Galilean
invariance \cite{Engel}. Note that we did not write the spin-orbit
interaction since it is absent in spin-unpolarized matter.

In uniform matter, the functional~(\ref{eq:skyrmeedf}) gives rise to a
constant Hartree-Fock (HF) potential $U_{\mathrm{HF}}$ and an
effective mass $m^*$. The former is the first derivative of
\Eq{eq:skyrmeedf} with respect to $\rho$, while the effective
mass is due to the $\tau$ dependence of the Skyrme functional
\cite{ChabanatSLy4}:
\begin{subequations}
   \begin{align}
      U_{\mathrm{HF}} &= \frac{s_0}{2} \rho 
        + \frac{\alpha+2}{24} s_3 \, \rho^{\alpha + 1} 
        + \frac{s_1 + 3 s_2}{8} \tau \, ,
      \label{eq:uhf}\\
      \frac{1}{2 \, m^*} &= \frac{1}{2m} + \frac{s_1 + 3 s_2}{8} \rho \, .
      \label{eq:masseff}
   \end{align}
\end{subequations}
We absorb $U_{\mathrm{HF}}$ in an effective chemical potential $\mu^*
= \mu - U_\HF$, so that the single-particle spectrum can be written as
   \begin{equation}
      \xi_{\vk} = \epsilon_{\vk} - \mu = \frac{k^2}{2m^*} - \mu^* \,.
      \label{eq:xik}
   \end{equation}

To study collective excitations within the RPA (or QRPA), one needs
the residual interaction between quasiparticles. The corresponding
matrix elements in the ph channel are obtained from the Skyrme
functional as follows \cite{Garcia}:
   \begin{align}
      V^\ph_{\vk_1,\vk_2,\vk_4,\vk_3} 
        = \frac{\delta^2 E_\Skyrme}
           {\delta\rho_{\vk_1,\vk_2}\,\delta\rho_{\vk_4,\vk_3}} \, ,
      \label{eq:vphdefinition}
   \end{align}
where $E_\Skyrme = \int d^3 r \, \mathcal{E}_\Skyrme$ is the energy.
The conservation of the total momentum $\vek{q}$ of the ph pair
implies that $V^{\ph}$ is proportional to
$\delta_{\vk_1-\vk_2,\vk_3-\vk_4}$. After transformation to relative
and total momenta of the ph pairs, the matrix element can conveniently
be written in the form \cite{Navarro}
   \begin{multline}
      V^\ph_{\vk+\frac{\vq}{2},\vk-\frac{\vq}{2},\vkp-\frac{\vqp}{2},\vkp+\frac{\vqp}{2}}
        = [W_1(q) + W_2\, (\vk-\vkp)^2] \delta_{\vq,\vqp} \,.
      \label{eq:vphexpression}
   \end{multline}
The explicit expressions for $W_1(q)$ and $W_2$ in terms of the
parameters of the Skyrme functional are given in the
Appendix~\ref{apx:skyrmeParameters}.
\subsection{Pairing interaction}
In order to account for the superfluidity of the neutron gas, we have
to include pairing. We do this in the framework of the
Bardeen-Cooper-Schrieffer (BCS) theory \cite{BCS}. Here, we consider
only pairing in the $^1S_0$ channel, i.e., of neutrons with opposite
spins, and disregard the $^3P_2$ channel, which becomes dominant at
higher densities \cite{Tamagaki}. If we define the anomalous density
by
\begin{equation}
\kappa_{\vk,\vkp} = \mean{a_{-\vkp\downarrow}a_{\vk\uparrow}}\,,
\label{eq:kappaOperator}
\end{equation}
the pairing gap $\Delta$ is given by the gap equation
\begin{equation}
\Delta_{\vk_1,\vk_2} = - \sum_{\vk_3,\vk_4}
    V^\pp_{\vk_1,\vk_2,\vk_4,\vk_3} \kappa^{\phantom{pp}}_{\vk_3,\vk_4} \, ,
    \label{eq:pairingmf}
\end{equation}
where $V^\pp_{\vk_1,\vk_2,\vk_4,\vk_3}$ is the matrix element of the
pairing interaction (for outgoing particles $\vk_1\uparrow$ and
$-\vk_2\downarrow$, and incoming particles $\vk_3\uparrow$ and
$-\vk_4\downarrow$).

In nuclear structure calculations with Skyrme interaction, usually a
contact interaction with (possibly) density dependent coupling
constant and a cut-off is employed (see
e.g. \cite{KhanSandulescu2002}). Here, we take a different approach
and use a simple separable approximation to a low-momentum interaction
$(V_\lowk)$ derived from a realistic nucleon-nucleon force
\cite{Bogner2007}. This interaction gives a reasonable density
dependence of the superfluid critical temperature in low-density
neutron matter \cite{RamananUrban}. The approximation we use is
   \begin{multline}
      V^{\pp}_{\vk_1,\vk_2,\vk_4,\vk_3} 
        = -g F(\tfrac{1}{2}|\vk_1+\vk_2|)  \, F(\tfrac{1}{2}|\vk_3+\vk_4|)\\
          \times \delta_{\vk_1-\vk_2,\vk_3-\vk_4}\,,
      \label{eq:vpp}
   \end{multline}
where $g$ is the strength of the interaction and $F$ is a Gaussian
form factor
   \begin{equation}
      F(k) = e^{-k^2/k_0^2} \,.
      \label{eq:formfactor}
   \end{equation}

In the ground state, $\kappa$ and $\Delta$ are diagonal, and we define
$\Delta_{\vk} = \Delta_{\vk,\vk}$. Then the gap equation reads
   \begin{align}
      \Delta_{\vk} = -\sum_{\vkp} V^{pp}_{\vk,\vk,\vkp,\vkp} \,
      \frac{\Delta_{\vkp}}{2 E_{\vkp}} \, ,
      \label{eq:gapeq}
   \end{align}
with the usual quasiparticle energy
   \begin{align}
      E_{\vk} = \sqrt{\xi_{\vk}^2 + \Delta_{\vk}^2} \,.
      \label{eq:quasipenergy}
   \end{align}
The separable form of the pairing interaction simplifies a lot the
solution of the gap equation: it is evident that $\Delta_\vk$ is
of the form $\Delta_{\vk}=\Delta_0 F(k)$, and instead of an
integral equation for the function $\Delta_\vk$ one has to solve
only an equation for the number $\Delta_0$.
\subsection{Quasiparticle Random Phase Approximation}
The QRPA treats small oscillations around the Hartree-Fock-Bogoliubov
(HFB) ground state (which, in the case of uniform matter, is obtained
by combining the HF and BCS frameworks discussed in the preceding
subsections) \cite{RingSchuck}.  It can be derived by linearising the
time dependent HFB (TDHFB) equations, see, e.g.,
\Ref{KhanSandulescu2002}, or, equivalently, by using the formalism of
normal and anomalous Green's functions, see, e.g.,
\Refs{SedrakianKeller2010,BaldoDucoin2011}. Here we use the TDHFB
formalism.

In addition to normal and anomalous density matrices $\rho$ and
$\kappa$ defined in \Eqs{eq:rhoOperator} and (\ref{eq:kappaOperator}),
we define
\begin{align}
\bar{\rho}_{\vk,\vkp} &=
  \mean{a^\dag_{-\vk\downarrow} a_{-\vkp\downarrow}^{\phantom{\dag}}}\,, &
\kappa^\dag_{\vk,\vkp} &=
  \mean{a^\dag_{\vkp\uparrow} a^\dag_{-\vk\downarrow}}\,.
\end{align}
Then the TDHFB equations can conveniently be written as \cite{RingSchuck}
\begin{equation}
  i \dot{\calR} = [\calH,\calR] \,,\label{eq:tdhfbeom}
\end{equation}
 with
\begin{align}
  \calH &=
    \begin{pmatrix}
      h&\Delta\\
      \Delta^\dag&-\bar{h}
    \end{pmatrix} \, , &
  \calR &= 
    \begin{pmatrix}
      \rho&-\kappa\\
      -\kappa^\dag&1-\bar{\rho}
    \end{pmatrix} \, .
\end{align}
The matrices $h$ and $\bar{h}$ denote the matrices of the one-body
mean-field hamiltonian which will be specified below.

As mentioned before, the QRPA is the linearization of the TDHFB
equations for small oscillations around the ground state. We therefore
split the matrices $\calR$ and $\calH$ into their ground-state values
$\Rz$ and $\Hz$ and small deviations $\Ru$ and $\Hu$. Let us first
look at the ground state, which of course has to satisfy
\Eq{eq:tdhfbeom} with $\dRz = 0$. This is the case because $\Hz$ and
$\Rz$ can be simultaneously diagonalized. In the ground state, we have
$h^{(0)}_{\vk,\vkp}=\bar{h}^{(0)}_{\vk,\vkp} =
\xi_{\vk}\delta_{\vk,\vkp}$ and $\Delta^{(0)}_{\vk,\vkp}
=\Delta^{\dag(0)}_{\vk,\vkp} =\Delta_{\vk}\delta_{\vk,\vkp}$, and the
matrix $\calH^{(0)}$ is diagonalized by the transformation
\begin{equation}
\tHz = \calW^T \Hz\calW = 
  \begin{pmatrix}E& 0\\ 0& -E\end{pmatrix}\,,
\label{eq:diago}
\end{equation}
with the eigenvalues $E_{\vk,\vkp} =
E_{\vk} \, \delta_{\vk,\vkp}$ and the transformation matrix
\begin{equation}
\calW =
  \begin{pmatrix} u & -v \\ v & u \end{pmatrix} \,,
\end{equation}
where $u$ and $v$ are the usual factors appearing in BCS theory
\begin{align}
  u_\vk &= \sqrt{\frac{1}{2}+\frac{\xi_\vk}{2E_\vk}}\,,&
  v_\vk &= \sqrt{\frac{1}{2}-\frac{\xi_\vk}{2E_\vk}}\,.
  \label{eq:ukvkdef}
\end{align}
The normal and anomalous density matrices in the ground state are
given by $\rho^{(0)}_{\vk,\vkp} = \bar{\rho}^{(0)}_{\vk,\vkp} =
v_{\vk}^2 \delta_{\vk,\vkp}$ and $\kappa^{(0)}_{\vk,\vkp} =
\kappa^{\dag(0)}_{\vk,\vkp} = u_{\vk} v_{\vk} \delta_{\vk,\vkp}$, so
that the same transformation diagonalizes $\calR^{(0)}$, too:
\begin{equation}
\tRz = \calW^T \Rz\calW = 
  \begin{pmatrix}0& 0\\ 0& 1\end{pmatrix}\,.
\end{equation}

Let us now consider a small perturbation of the system. By keeping in
\Eq{eq:tdhfbeom} only the first order in the deviations, we
obtain the linearized equation of motion
\begin{equation}
i \dRu = [\Hz,\Ru]+[\Hu,\Rz]\,.
\end{equation}
The equation can be simplified by applying again the transformation
that diagonalizes $\Hz$ and $\Rz$. After a Fourier transform with
respect to time one obtains the following equation:
\begin{equation}
  \label{matrix_diag_R}
  \omega \tRu = 
    \begin{pmatrix}
       [E,\tRu_{11}]&\{E,\tRu_{12}\} + \tHu_{12} \\
       -\{E,\tRu_{21}\} - \tHu_{21}&-[E,\tRu_{22}]
    \end{pmatrix} \,,
\end{equation}
from which one can easily determine the non-vanishing elements
$\tRu_{12}$ and $\tRu_{21}$ as functions of $\tHu_{12}$ and
$\tHu_{21}$. The matrix $\Ru$ is then obtained by transforming $\tRu$
back. The resulting expressions are lengthy, but they can be
simplified by using the following linear combinations:
\begin{align}
   \rho^\pm &= \rho \pm \bar{\rho} \,, &
   \kappa^\pm &= \kappa \pm \kappa^{\dag} \,, \\
   h^\pm &= h \pm \bar{h} \,, &
   \Delta^{\pm} &= \Delta \pm \Delta^\dag \,.
\end{align}
In the case of spin-independent excitations studied in the present
paper, $\rho^+$ is responsible for density oscillations, while
$\rho^-$ describes the corresponding current. (In the case of spin
modes, the situation would be reversed.) The quantities $\kappa^+$ and
$\Delta^+$ are related to oscillations of the amplitude of $\Delta$,
while $\kappa^-$ and $\Delta^-$ describe phase oscillations which are
extremely important in the context of the low-energy collective mode
(Goldstone mode). The solution for $\rho^{\pm(1)}$ and
$\kappa^{\pm(1)}$ can be written in the form
\begin{equation}
  \begin{pmatrix}
         &\rho^{+(1)}_{\vk_1,\vk_2} \\
         &\rho^{-(1)}_{\vk_1,\vk_2} \\
         &\kappa^{+(1)}_{\vk_1,\vk_2} \\
         &\kappa^{-(1)}_{\vk_1,\vk_2} 
      \end{pmatrix}
      = \Pi^{(0)}_{\vk_1,\vk_2}(\omega)
      \begin{pmatrix}
         &h^{+(1)}_{\vk_1,\vk_2} \\
         &h^{-(1)}_{\vk_1,\vk_2} \\
         &\Delta^{+(1)}_{\vk_1,\vk_2} \\
         &\Delta^{-(1)}_{\vk_1,\vk_2} 
      \end{pmatrix}
      \label{eq:respfuncQRPA}
\end{equation}
where $\Pi^{(0)}_{\vk_1,\vk_2}(\omega)$ is a $4\times 4$ matrix whose
components denoted by $\Pi^{\rho^+,h^+}_{\vk_1,\vk_2}$, \dots,
$\Pi^{\kappa^-,\Delta^-}_{\vk_1,\vk_2}$ are given in
Appendix~\ref{apx:RFMatrix}.

So far, we have not specified the perturbation of the hamiltonian,
$h^{(1)}$. There are two contributions of different origin. First, to
probe the system, we apply an external perturbation at $t=0$ of the
form of a plane wave, i.e., $V_\ex e^{i\vq\cdot\vek{r}} \delta(t)$, which
after Fourier transformation becomes $V_\ex
\delta_{\vk_1-\vk_2,\vq}$. The second contribution to $h^{(1)}$ comes
from the oscillations of the mean field due to the density
oscillations~:
\begin{equation}
  h^{(1)}_{\vk_1,\vk_2} = V_\ex \delta_{\vk_1-\vk_2,\vq} + 
    \sum_{\vk_3,\vk_4} V^\ph_{\vk_1,\vk_2,\vk_4,\vk_3} \, 
    \rho^{(1)}_{\vk_3,\vk_4}\,.
  \label{eq:phchannel}
\end{equation}
Analogously, the oscillation of the gap, $\Delta^{(1)}$, is related to
the oscillation of the anomalous density,
\begin{equation}
  \Delta^{(1)}_{\vk_1,\vk_2} = - \sum_{\vk_3,\vk_4} 
  V^\pp_{\vk_1,\vk_2,\vk_4,\vk_3} \, 
  \kappa^{(1)}_{\vk_3,\vk_4} \, .
  \label{eq:ppchannel}
\end{equation}
Looking at \Eq{eq:respfuncQRPA} and taking into account the
momentum conservation in the interactions $V^\ph$ and $V^\pp$, one
sees that an external perturbation proportional to
$\delta_{\vk_1-\vk_2,\vq}$ leads to non-vanishing elements of
$\rho_{\vk_1,\vk_2}$ and $\kappa_{\vk_1,\vk_2}$ only for $\vk_1-\vk_2=\vq$. This
could have been anticipated, since in a uniform system a perturbation
having the form of a plane wave can only excite oscillations which are
also plane waves with the same wave vector as the perturbation. We
therefore introduce the short-hand notation $\vk_{\pm} =
\vk\pm\frac{\vq}{2}$ and denote the non-vanishing matrix elements by
$\rho_{\vk_+,\vk_-}$, etc.

The advantage of the Skyrme functional is that $h^{(1)}$ depends only on local
quantities. With the notation of \Eq{eq:vphexpression}, we have
\begin{subequations}
\begin{align}
  h^{+(1)}_{\vk_+,\vk_-} &= W_1(q)\, \rho^{+(1)}_{\vek{q}}
    + W_2 k^2\, \rho^{+(1)}_{\vek{q}} + W_2\, \tau^{+(1)}_{\vek{q}} + 2V_\ex\,, \\
  h^{-(1)}_{\vk_+,\vk_-} &= 2 W_2 k \cos\theta\, j^{-(1)}_{\vek{q}} \, ,
\end{align}
\label{eq:vphoe}
\end{subequations}
where $\theta$ is the angle between $\vk$ and $\vq$ and
\begin{subequations}
   \begin{align}
      \rho^{+(1)}_{\vek{q}} &= \sum_{\vk} \rho^{+(1)}_{\vk_+,\vk_-} \, , \\
      \tau^{+(1)}_{\vek{q}} &= \sum_{\vk} k^2 \, \rho^{+(1)}_{\vk_+,\vk_-} \, , \\
      j^{-(1)}_{\vek{q}} &= \sum_\vk k \, \cos \theta \, \rho^{-(1)}_{\vk_+,\vk_-} \, .
   \end{align}
   \label{eq:rhoFamilyDef}
\end{subequations}
Similarly, in the pp channel, the calculation is simplified by the
fact that our pairing interaction (\ref{eq:vpp}) is separable:
\begin{equation}
  \Delta^{\pm(1)}_{\vk_+,\vk_-} = g F(k) \kappa_\vq^{\pm(1)}
\label{eq:vppoe}
\end{equation}
with
\begin{equation}
  \kappa_\vq^{\pm(1)} = \sum_\vk F(k) \kappa^{\pm(1)}_{\vk_+,\vk_-}
\label{eq:kappaFamilyDef}
\end{equation}

Now we are able to calculate the linear response by inserting
\Eqs{eq:respfuncQRPA}, (\ref{eq:vphoe}) and (\ref{eq:vppoe}) into
\Eqs{eq:rhoFamilyDef} and (\ref{eq:kappaFamilyDef}). In this way we obtain
   \begin{align}
      \begin{pmatrix}
         \rho^{+(1)}_{\vek{q}} \\
         \tau^{+(1)}_{\vek{q}} \\
         j^{-(1)}_{\vek{q}} \\
         \kappa^{+(1)}_{\vek{q}} \\
         \kappa^{-(1)}_{\vek{q}}
      \end{pmatrix}
      =
      \bigg( \mathbb{I} - \sumk{\Pi^{(0)}_{\vek{q}}V} \bigg)^{-1}
      \begin{pmatrix}
         \sumk{\Pi^{\rho+,h+}_{\vk_+,\vk_-}} \\
         \sumk{k^2 \, \Pi^{\rho^+,h^+}_{\vk_+,\vk_-}} \\
         \sumk{k \cos \theta \, \Pi^{\rho^-,h^+}_{\vk_+,\vk_-}} \\
         \sumk{F(k) \Pi^{\kappa^+,h^+}_{\vk_+,\vk_-}} \\
         \sumk{F(k) \Pi^{\kappa^-,h^+}_{\vk_+,\vk_-}} 
      \end{pmatrix}
      2V_\ex \, ,
      \label{eq:qrpaRF}
   \end{align}
   where the short-hand notation $\sumk{f(\vk)}$ denotes the sum of
   $f(\vk)$ over $\vk$,
   \begin{align}
      \sumk{f(\vk)} = \sum_{\vk} f(\vk) \,,
      \label{eq:sumOverK}
   \end{align}
and the matrix $\sumk{\Pi^{(0)}_{\vek{q}}V}$ is given in
Appendix~\ref{apx:RFMatrix}. \\

It is well known that superfluidity leads to the existence of the
so-called Bogoliubov-Anderson sound \cite{Bogoliubov,Anderson}, a
collective mode with linear dispersion relation $\omega \propto q$
(for small $q$) which can be interpreted as a Goldstone boson
corresponding to the broken $\rm{U}(1)$ symmetry
\cite{Weinberg2}. This implies that the QRPA response function has a
pole at low energy. The energy $\omega$ of this collective mode can be
found by searching for a given $q$ the root of the determinant of the
matrix appearing in \Eq{eq:qrpaRF}:
   \begin{align}
      \left | \mathbb{I} - \sumk{\Pi^{(0)}_{\vek{q}}V} \right | = 0 \, .
      \label{eq:detRoots}
   \end{align}
This collective mode exists only at low momentum $q$, as long as its
energy $\omega$ lies below the pair-breaking threshold $\sim
2\Delta_{k_F}$, where $k_F$ denotes the Fermi momentum. At higher
values of $q$, the collective mode enters the two-quasiparticle
continuum and gets a width (finite lifetime).
\subsection{Landau approximation}
\label{sec:Formalism:Landau}
In some recent work \cite{BaldoDucoin2011,Keller}, the QRPA response
was calculated within the Landau approximation
\cite{Nozieres_english}. In this approximation, one exploits the fact
that for small $q$ the change of the density matrix $\rho_{\vk,\vkp}$
is concentrated at the Fermi surface, $|\vk|\approx|\vkp|\approx
k_F$. Keeping only the Landau parameter $F_0$ amounts to replacing
\Eq{eq:vphoe} by
\begin{equation}
h^{+(1)}_{\vk_+,\vk_-} = (W_1(0)+2W_2k_F^2) \rho^{+(1)}_{\vek{q}}
\end{equation}
and neglecting $h^{-(1)}_{\vk_+,\vk_-}$. However, because of the
effective mass $m^* \neq m$, this approximation violates Galilean
invariance \cite{Nozieres_english} and one should also include the
parameter $F_1$. In this case one has
\begin{equation}
h^{-(1)}_{\vk_+,\vk_-} = 2 W_2 k_F \cos\theta\, j^{-(1)}_{\vek{q}}\,,
\end{equation}
where the current $j^{-(1)}_{\vek{q}}$ is calculated from
\begin{equation}
j^{-(1)}_{\vek{q}} = k_F\sum_\vk \cos \theta \, \rho^{-(1)}_{\vk_+,\vk_-} \, .
\end{equation}
As a consequence, the $5\times 5$ matrix in \Eq{eq:qrpaRF} reduces to
a $3\times 3$ or $4\times 4$ one if one keeps only $F_0$ or $F_0$ and
$F_1$, respectively.
\subsection{Hydrodynamics}
The famous result for the dispersion relation of the
Bogoliubov-Anderson mode, $\omega = k_F q/(\sqrt{3}m)$, first derived
by Bogoliubov \cite{Bogoliubov} and Anderson \cite{Anderson}, would be
correct in an ideal Fermi gas. Leggett \cite{Leggett_ColMod}
generalized this result in the framework of Landau's Fermi-liquid
theory to include the interaction among quasiparticles. In both cases
the sound velocity $u = \omega/q$ agrees with the hydrodynamic one,
\begin{equation}
u^2 = \left. \frac{1}{m\rho} \frac{\partial P}{\partial \rho} \right |_s
\end{equation}
($P$ and $s$ are the pressure and the entropy density, respectively),
which in the zero-temperature case can be simplified to
\begin{equation}
u^2 = \left. \frac{1}{m} \frac{\partial \mu}{\partial \rho} \right |_{T=0} \,,
\label{eq:hydrosound}
\end{equation}
since $s=0$ at $T=0$.

At a first glance, it is surprising that hydrodynamics is applicable
here. In a normal fluid, hydrodynamics requires collisions that
restore local equilibrium. Otherwise, in the collisionless regime, the
local Fermi sphere gets deformed during the oscillation, which gives
rise to the so-called zero-sound modes \cite{Nozieres_english}. The
situation is completely different in a superfluid at $T=0$: although
there are no collisions, the local Fermi sphere stays spherical during
the oscillation because of pairing. This ``superfluid hydrodynamics''
was also used to describe collective modes in trapped (i.e.,
non-uniform) Fermi gases \cite{Menotti2002}, and in
Ref. \cite{Grasso2005} it was demonstrated that also in that case
hydrodynamic and QRPA results for $T=0$ agree if pairing is strong
enough.

In order to calculate the hydrodynamic speed of sound, we use in
\Eq{eq:hydrosound} the chemical potential obtained with the Skyrme
functional (with pairing).
\section{Results}
\label{sec:Results}
\subsection{Ground state}
   \label{para:gapeq_resolution}
Before we turn to the linear response, let us briefly discuss the
ground state properties. For the mean field, we use the SLy4
parametrization of the Skyrme force, whose parameters are given in
\Ref{ChabanatSLy4}. This interaction was not only fitted to nuclei,
but also to the EOS of neutron matter. Since pairing has only a
marginal effect on the EOS, our EOS agrees with that shown, e.g., in
\Ref{DouchinHaensel2000}.

\begin{figure}
  \includegraphics[width=8cm]{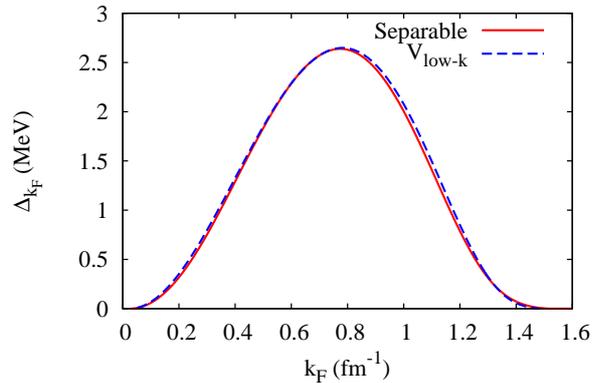}
\caption{(Color online) Value of the gap at the Fermi surface,
  $\Delta_{k_F}$, as function of the Fermi momentum $k_F$, obtained
  with the separable interaction (solid line) and with the $V_{\lowk}$
  interaction of \cite{Bogner2007} (dashes).}
  \label{fig:deltaKf}
\end{figure}

To determine the two parameters $g$ and $k_0$ of our pairing
interaction, \Eqs{eq:vpp} and (\ref{eq:formfactor}), we first solve
the gap equation (\ref{eq:gapeq}) with the (non-separable) $V_{\lowk}$
interaction\footnote{The matrix elements used here are those obtained
  in \Ref{Bogner2007} with a Fermi-Dirac regulator with $\Lambda =
  2$~fm$^{-1}$ and $\epsilon = 0.5$~fm$ ^{-1}$.}. The resulting gap at
the Fermi surface, $\Delta_{k_F}$, as a function of $k_F =
(3\pi^2\rho)^{1/3}$, is displayed in \Fig{fig:deltaKf}
(dashes). Then we fit $g$ and $k_0$ to reproduce this result
with the separable interaction. The result of this fit is also shown
in \Fig{fig:deltaKf} (solid line), and the corresponding parameter
values are listed in Table~\ref{tab:vppparam}.
\begin{table}
\caption{Parameters of the pairing interaction, \Eqs{eq:vpp} and
  (\ref{eq:formfactor}).\label{tab:vppparam}}
   \begin{ruledtabular}
   \begin{tabular}{lr}
      $g$ (MeV fm$^{3}$) & 856 \\
      $k_0$ (fm$^{-1}$) & 1.367\\
   \end{tabular}
   \end{ruledtabular}
\end{table}
We see that with this pairing interaction, the maximum of the gap,
$\Delta_{k_F} \sim 2.7$~MeV, is reached at $k_F \sim 0.8$~fm$^{-1}$,
corresponding to a density of $\rho \sim 0.017$~fm$^{-3}$. At low
density, the gap increases with density because of the increasing
level density at the Fermi surface. The decrease of the gap at high
density is due to the form factor, \Eq{eq:formfactor}, and not due to
an explicit density dependence of the pairing interaction as it is
often used in HFB and QRPA calculations with Skyrme forces (see, e.g.,
\Ref{KhanSandulescu2002}). The fact that our maximum gap is reduced by
$\sim 10\,\%$ compared to typical BCS results obtained with the free
nucleon mass \cite{Hebeler2007} is a consequence of the reduction of the
density of states due to the effective mass $m^* < m$.

However, it should be pointed out that there is no consensus in the
literature about the correct density dependence of the gap
\cite{Dean2003,Chamel2012}, mainly because of screening effects beyond
BCS theory (analogous to the Gorkov--Melik-Barkhudarov correction
\cite{Gorkov1961}), which could lead to a dramatic suppression of the
gap. At low density, recent Quantum-Monte-Carlo calculations
\cite{Gezerlis2008,GezerlisCarlson2010} seem to be reliable and show a
suppression of the gap between $30$ and $50\,\%$ compared to the BCS
result.
\subsection{QRPA response function and collective mode}
We will now study the QRPA response function in neutron matter for
different densities and compare it with the RPA one. In the present
work we consider the density response, which is defined by
$\Pi(\omega,q) = \rho^{+(1)}_{\vq}/(2V_\ex)$. Since its real and
imaginary parts are related to each other via dispersion relations, it
is enough to discuss the imaginary part, the so-called strength
function.

\begin{figure}
\includegraphics[width=8cm]{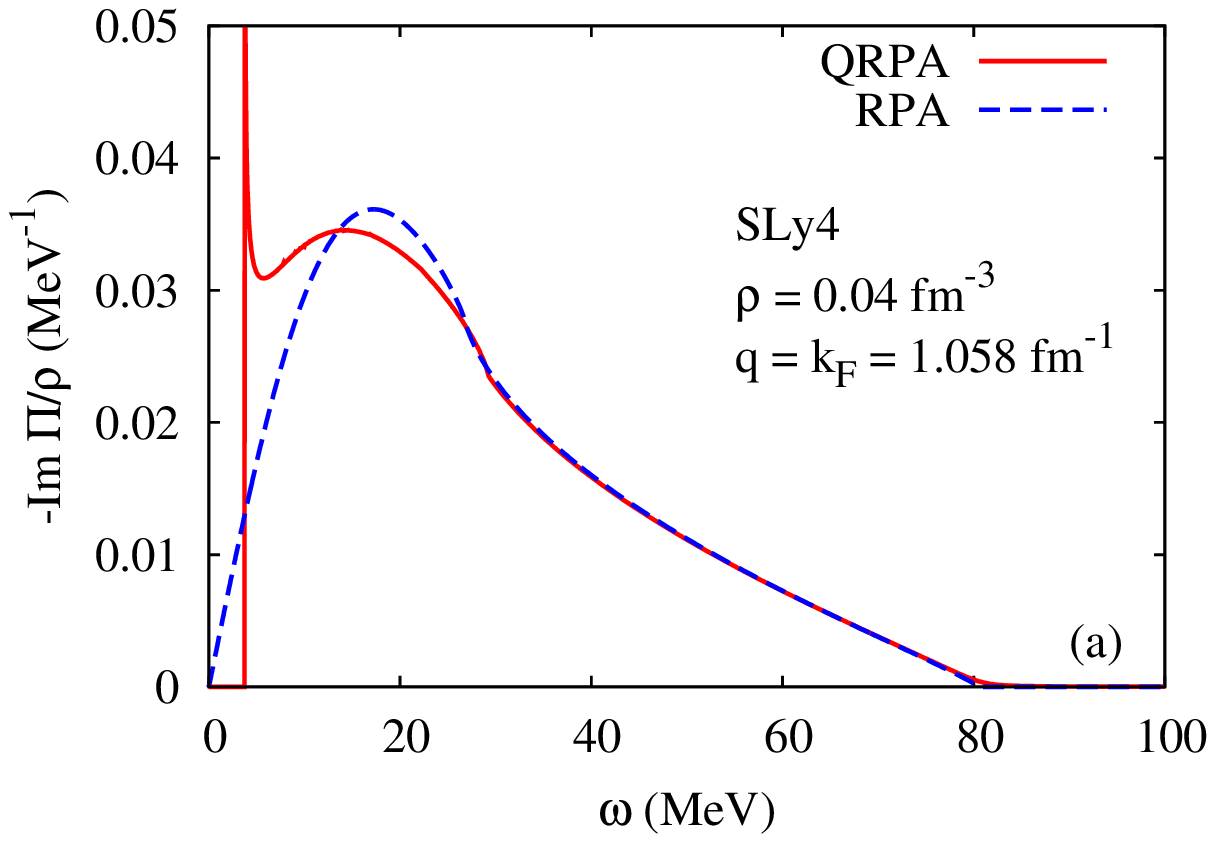}
\\
\includegraphics[width=8cm]{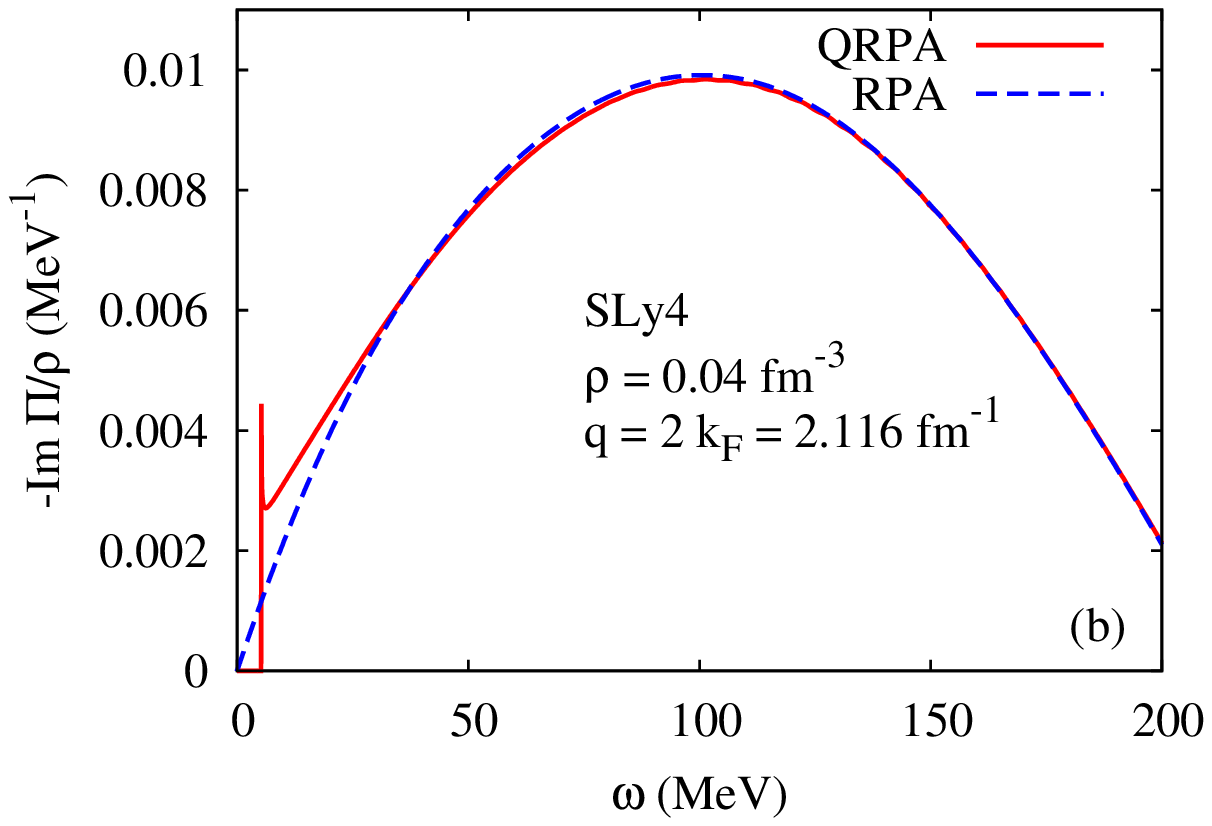}
\caption{(Color online) QRPA (solid lines) and RPA (dashes) response
  functions at density $\rho = 0.04$~fm$^{-3}$, as functions of the
  excitation energy $\omega$ for two different momentum transfers $q =
  k_F$ (a) and $2 k_F$ (b).
\label{fig:RPA_QRPA_RF}}
\end{figure}

We choose densities between $0.016$ and $0.04$~fm$^{-3}$,
corresponding to typical densities of the neutron gas surrounding the
clusters in the inner crust of a neutron star \cite{Negele}. At higher
densities, as they are realized in the neutron star core, our approach
is not valid because there the neutrons are paired in the $^3P_2$
channel \cite{Tamagaki}. As we have seen in the preceding subsection, the
$^1S_0$ gap decreases with increasing density. We therefore expect
that at high density, our QRPA response approaches the RPA one. The
latter is the response calculated without pairing, i.e., by setting
$\Delta_\vk = 0$ and keeping only the upper left $3\times3$ part of
the matrix in \Eq{eq:qrpaRF}, and we checked that it coincides with
the RPA response functions that can be found in the literature
\cite{Garcia}. As one can see in \Fig{fig:RPA_QRPA_RF},
where the strength function is shown for $\rho = 0.04$~fm$^{-3}$, the
RPA (dashes) and QRPA (solid lines) responses are indeed similar and
approach each other with increasing excitation energy $\omega$ and
momentum transfer $q$. For $q = k_F$ (upper panel) and $2 k_F$ (lower
panel), the RPA strength function has a broad continuum. The effect of
pairing is to shift the threshold of the continuum from zero to the
pair-breaking threshold $\sim 2\Delta_{k_F}$. At excitation energies
much larger than $2\Delta_{k_F}$, the response is practically not
affected by pairing. At energies around the threshold, however, the
response is strongly modified by pairing. The peak visible at the
threshold corresponds to a collective mode which is damped since it
lies in the continuum, i.e., it can decay into two quasiparticles. In
RPA, one does not see any collective mode, since the ph interaction is
attractive and a collective zero-sound mode, as it can be described by
RPA, exists only for repulsive ph interaction \cite{Nozieres_english}.

\begin{figure}
\includegraphics[height=8cm,angle=-90]{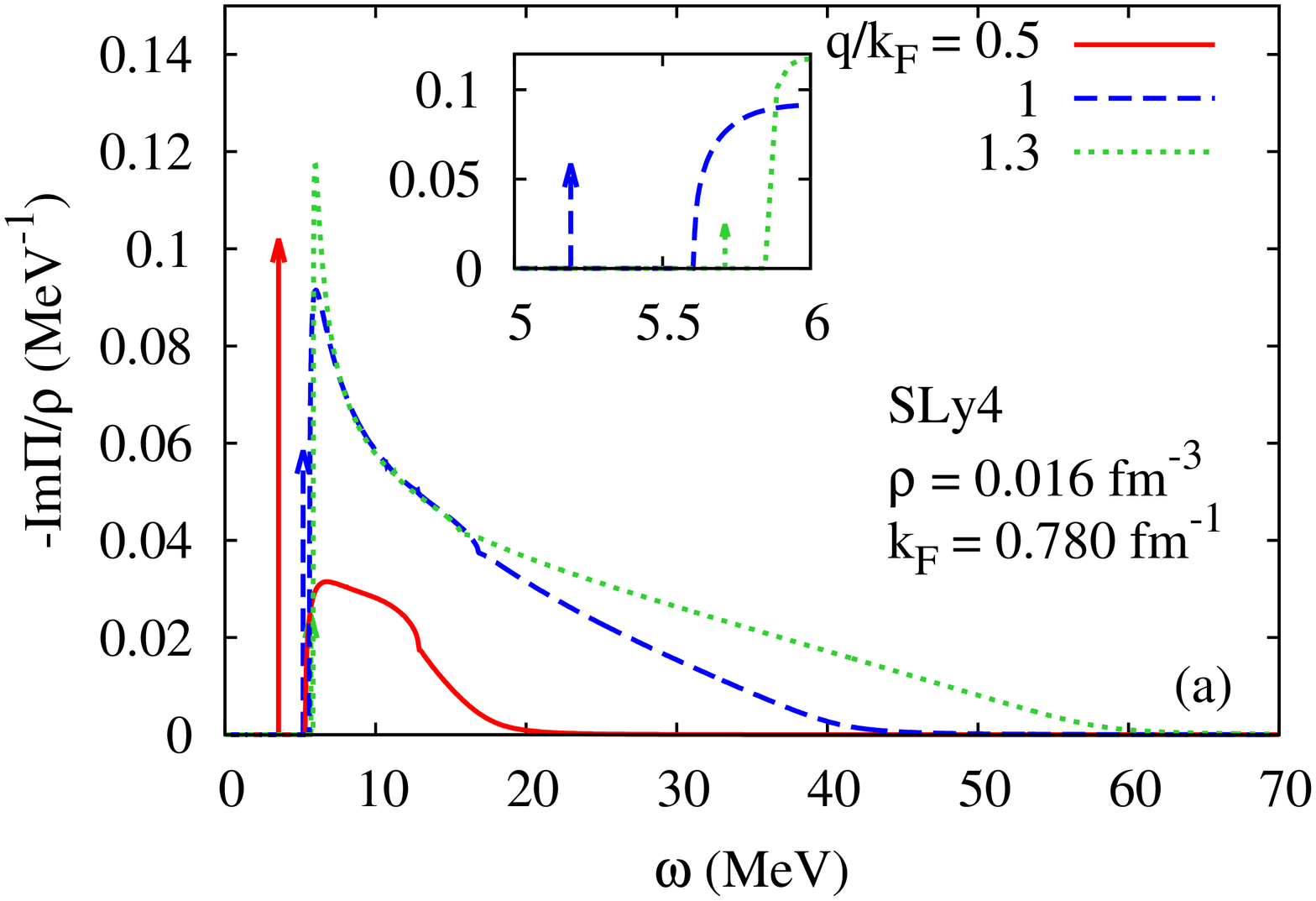}\\
\includegraphics[height=8cm,angle=-90]{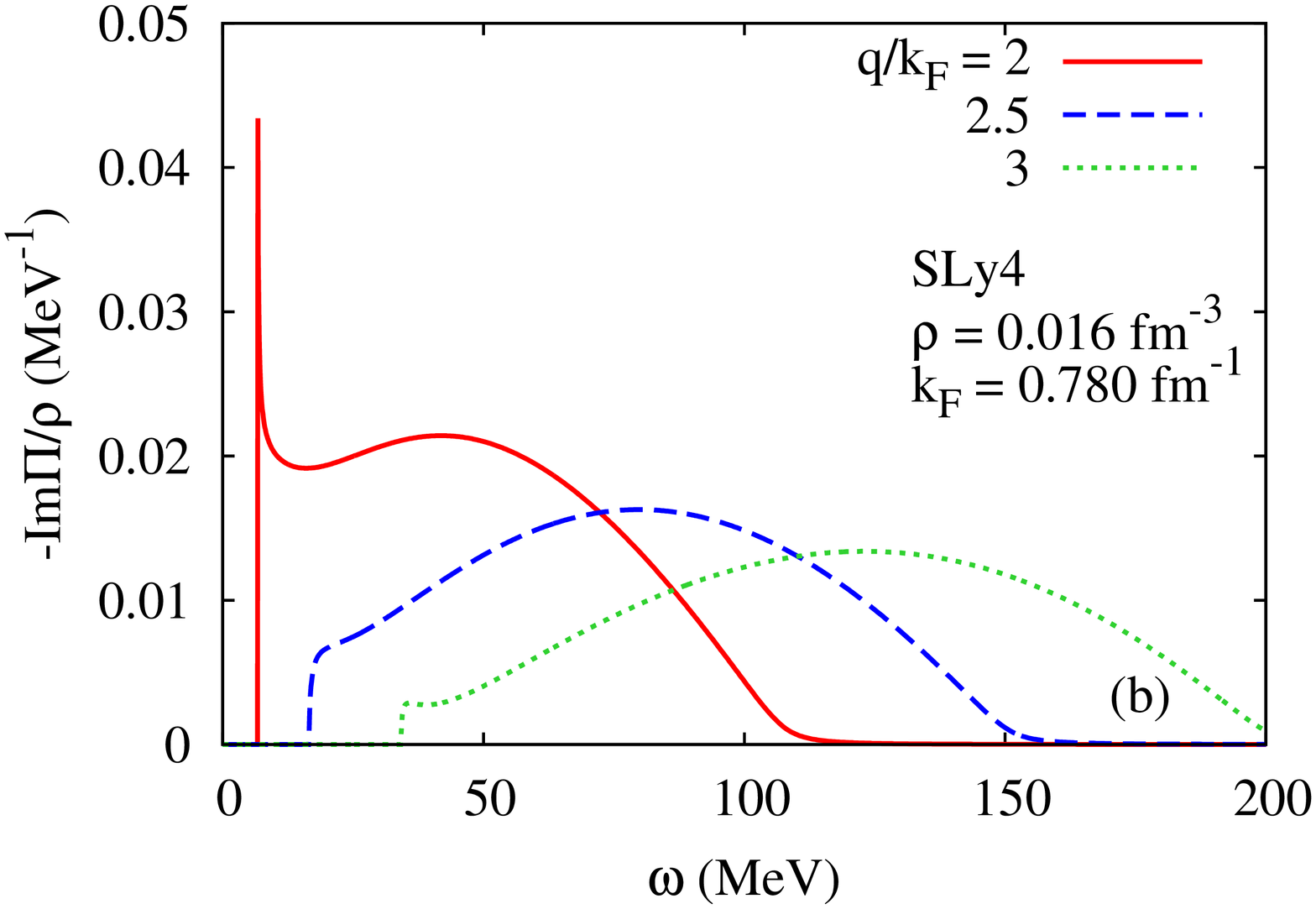}
\caption{(Color online) QRPA response functions for $q/k_F = 0.5, 1,
  1.3$ (a) and $2, 2.5, 3$ (b) at $\rho =
  0.016$~fm$^{-3}$ as functions of the excitation energy $\omega$. The
  arrows in the upper panel represent $\delta$-function peaks
  corresponding to the undamped collective modes. Their height is
  proportional to their strength which corresponds to 71.3\%
  ($q=0.5\,k_F$), 25.2\% ($q = k_F$), and 9.5\% ($q = 1.3\, k_F$) of
  the total strength of the response function. In the lower panel (b), the
  collective mode lies above the continuum threshold.}
\label{fig:qrpa_rf}
\end{figure}

In the preceding examples the collective mode was damped because we
considered a high momentum $q$ and relatively weak pairing. In order
to see more clearly the collective mode, let us now choose a lower
density $\rho = 0.016$~fm$^{-3}$ and smaller momenta. In the upper
panel of \Fig{fig:qrpa_rf},
we see the imaginary part of the response function for momenta between
$q = 0.5 \, k_F$ and $1.3 \, k_F$. Now there is a pole in the real part of
the response function below the continuum threshold, corresponding to
an undamped collective mode. In principle, the imaginary part has a
$\delta$-function peak at this energy, which is represented as an
arrow in \Fig{fig:qrpa_rf}. The height of each arrow indicates the
strength contained in the peak, which is proportional to
the derivative $d(\Pi^{-1})/d\omega$ calculated at the pole of $\Pi$. We
can see that the strength is highest for small $q$ and decreases as
the mode approaches the continuum threshold. At momenta higher than
$\sim 1.5$~fm$^{-1}$ (see lower panel of \Fig{fig:qrpa_rf}), the
collective mode enters again into the continuum, as in
\Fig{fig:RPA_QRPA_RF}.

\begin{figure}
\includegraphics[width=8cm]{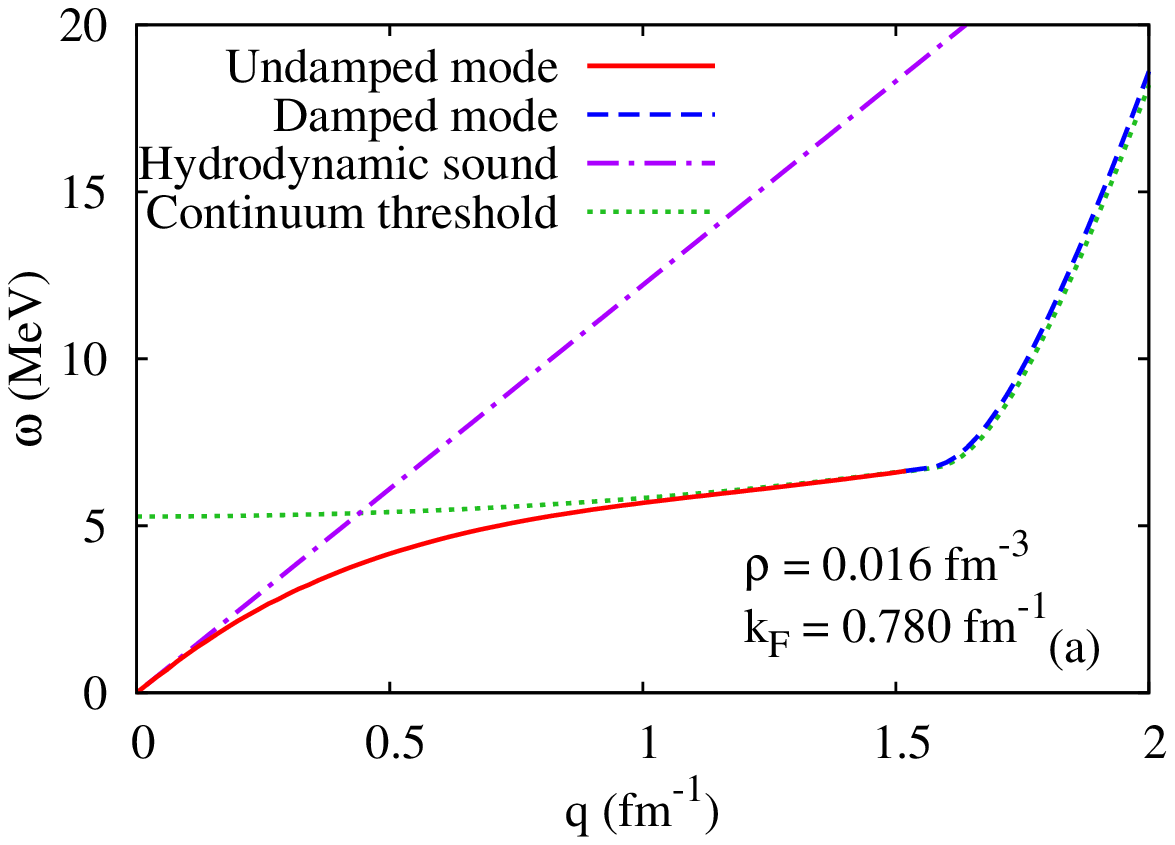}
\includegraphics[width=8cm]{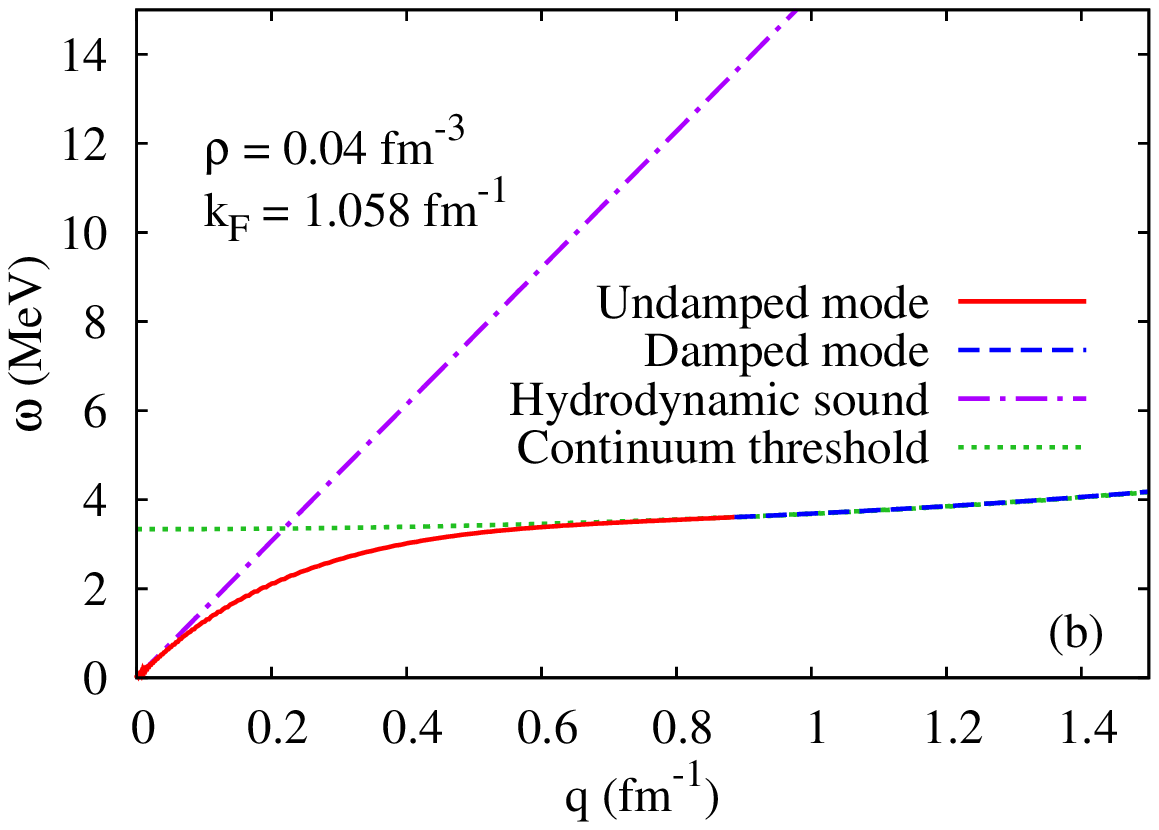}
\caption{(Color online) Dispersion relation $\omega_\vq$ of the
  undamped (solid line) and damped (dashes) collective mode at $\rho =
  0.016$ (a) and $0.04$~fm$^{-3}$ (b). At small
  $q$, it agrees with the linear dispersion relation $\omega = uq$ of
  hydrodynamic sound (dash-dotted line). At higher $q$, it approaches
  and finally crosses the pair-breaking threshold (dotted line).
\label{fig:colmod}}
\end{figure}

Let us study in more detail the dispersion relation $\omega_\vq$ of the
collective mode. In \Fig{fig:colmod},
the solid lines represent the dispersion relations of the undamped
collective mode at densities $\rho = 0.016$ (upper panel) and
$0.04$~fm$^{-3}$ (lower panel). We see that at small $q$, the
dispersion relation is practically linear. The fact that $\omega\to 0$
for $q\to 0$, as required by the Goldstone theorem, is in practice a
very good test of our numerics, since $\omega_{\vq=0}$ is extremely
sensitive to small numerical errors in the matrix
$\sumk{\Pi_\vq^{(0)}V}$. Another test is the slope $d\omega/dq$ at
$q=0$, which agrees perfectly with the hydrodynamic speed of sound
calculated from \Eq{eq:hydrosound} (dash-dotted lines). We see that
$\omega_\vq$ stays more or less linear as long as $\omega \ll
2\Delta_{k_F}$. Since in the case $\rho = 0.04$~fm$^{-3}$ the gap
$\Delta_{k_F}$ is smaller and the speed of sound $u$ is higher, the
range of applicability of the hydrodynamic approximation is smaller
than in the case $\rho = 0.016$~fm$^{-3}$. At larger $q$, the mode
frequency starts to bend and approaches the pair-breaking threshold,
which is represented by the dots (approaching $2\Delta_{k_F}$ and
$q(q/2-k_F)/m^*$, respectively, in the limits of very small and very
large $q/k_F$). Above a certain $q$, (e.g., $\sim 1.5$~fm$^{-1}$ for
$\rho = 0.016$~fm$^{-3}$ and $\sim 0.9$~fm$^{-1}$ for $\rho =
0.04$~fm$^{-3}$) the mode enters into the continuum (dashes) but it
stays practically at the threshold (cf. also lower panel of
\Fig{fig:qrpa_rf}). This behavior of the collective mode is
qualitatively different from the one shown in \Ref{Keller} but similar
to the one obtained in \Ref{BaldoDucoin2011}. Also in the context of
ultracold atoms, results similar to ours have been found, see
\Ref{Combescot} for a QRPA calculation and \Ref{Forbes2013} where the
collective mode was studied as small-amplitude oscillation in a
time-dependent density-functional theory implementation (similar to
TDHFB).
\subsection{Comparison with the Landau approximation}
\begin{figure}
      \includegraphics[width=8.5cm]{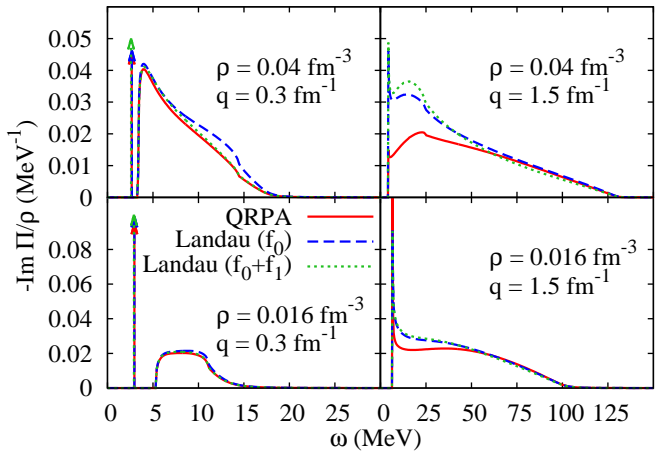}
   \caption{(Color online) Response functions obtained within the full
     QRPA (solid lines) and within the Landau approximation including
     only $F_0$ (dotted lines) or $F_0$ and $F_1$ (dashes) as
     functions of the excitation energy $\omega$ for neutron densities
     $\rho = 0.016$~fm$^{-3}$ (lower panels) and $0.04$~fm$^{-3}$
     (upper panels) and momenta $q = 0.3$~fm$^{-1}$ (left panels) and
     $1.5$~fm$^{-1}$ (right panels). The arrows in the left panels
     indicate positions and strengths of the collective modes. In
     QRPA, the strengths of the collective modes correspond to 82\%
     ($\rho = 0.016$~fm$^{-3}$) and 52\% ($\rho = 0.04$~fm$^{-3}$) of
     the total strength of the response functions.}
   \label{fig:landauApprox}
\end{figure}

Now we discuss the results obtained within the Landau approximation as
explained in Sec.~\ref{sec:Formalism:Landau}. This approximation has
recently been used in \Refs{BaldoDucoin2011,Keller}. In
\Fig{fig:landauApprox}
we display response functions for two different densities ($\rho =
0.016$ and $0.04$~fm$^{-3}$) and momenta ($q = 0.3$ and
$1.5$~fm$^{-1}$) within the Landau approximation keeping only $F_0$
(dotted lines), and within the Landau approximation keeping $F_0$ and
$F_1$ as required by Galilean invariance (dashes), and compare them
with the full QRPA results (solid lines). In the case of small
momentum transfer ($q = 0.3$~fm$^{-1}$, left panels of
\Fig{fig:landauApprox}), the three calculations give very similar
results. As in \Fig{fig:qrpa_rf}, the arrows indicate the energy and
strength of the undamped collective mode. We see that the Landau
approximation (with $F_0$ and $F_1$, and even with $F_0$ only) works
very well for the energy of the collective mode, only the strength
(height of the arrow) is slightly different from that obtained in the
full QRPA \footnote{It is well known that the sound velocity is given
  by $u^2 = k_F^2/(3m^{*\,2}) (1+F_0)(1+F_1/3)$
  \cite{Leggett_ColMod}. However, the last term depending on $F_1$
  does not originate from the residual interaction, but from the
  effective mass $m^*$, which is related to $F_1$ by Galilean
  invariance: $u^2 = k_F^2/(3mm^*) (1+F_0)$
  \cite{Nozieres_english}. Therefore, if one calculates the response
  function with the effective mass $m^*$, one already obtains the
  correct sound velocity by including only $F_0$ in the residual
  interaction.}. At excitation energies above $\sim 10$ MeV one starts
to see a difference between the two Landau approximations. As
expected, the result obtained with $F_0$ and $F_1$ is in better
agreement with the full QRPA than that obtained with $F_0$ only, as
one can see in the upper left panel of \Fig{fig:landauApprox}.

The situation is completely different at higher momentum transfer. In
the right panels of \Fig{fig:landauApprox}, we show results for $q =
1.5$~fm$^{-1}$. In this case, the collective mode has disappeared in
the continuum. Now the responses obtained within the Landau
approximation and within the full QRPA are clearly different. This is
not surprising, since the basic assumption underlying the Landau
approximation, namely that the excited quasiparticles are close to the
Fermi surface, is no longer fulfilled, and also the $q$ dependence of
the residual ph interaction [term $W_1(q)$] is no longer
negligible. We note that the inclusion of the $F_1$ Landau parameter
does not improve the agreement of the Landau approximation with the
full QRPA in this case.

To conclude, the Landau approximation seems to be sufficient to
establish the dispersion curve of the collective mode of the neutron
gas. However, it may strongly affect calculations that need the entire
response function, e.g. the neutrino mean free path in neutron stars
\cite{Margueron-Neutrino}.
\subsection{Heat capacity}
In \Ref{Fortin-Margueron} it was pointed out that neutron pairing
results in a strong suppression of the heat capacity at low
temperature, which might have observable effects on the neutron star
cooling. The relevant temperature range is $T\lesssim 10^9$~K $\sim
100$~keV, which is much smaller than $\Delta_{k_F}$ in the region we
are interested in. The quasiparticle contribution to the specific heat
at temperature $T$ can be obtained from
\begin{equation}
c_{v,\text{qp}} = T \, \left. \frac{\partial \, s_{\text{qp}}}{\partial T}
  \right |_{\rho} \,,
\label{eq:cvBCS}
\end{equation}
where $s_{\text{qp}}$ denotes the entropy density of thermally excited
quasiparticles \cite{Tinkham} 
\begin{multline}
s_{\text{qp}} = -2 \sum_\vp \big[\big(1-f(E_\vp)\big)\ln\big(1-f(E_\vp)\big)\\
  +f(E_\vp)\ln\big(f(E_\vp)\big)\big]
\end{multline}
with $f(E) = 1/(e^{E/T}+1)$. Indeed, $c_{v,\text{qp}}$ is suppressed
by a factor of $e^{-\Delta_{k_F}/T}$ at low temperature, as it is the
case in superconducting metals \cite{FetterWalecka}. Note that in a
superconductor, the Bogoliubov-Anderson mode is shifted upwards to the
plasma frequency by the Coulomb interaction \cite{Nambu1960} and
therefore its contribution to the specific heat is
negligible. However, in a superfluid such as the neutron gas the
situation is different because here the Bogoliubov-Anderson mode is
the dominant contribution to the specific heat at low temperature, and
not the quasiparticles.

\begin{figure}
  \includegraphics[width=8cm]{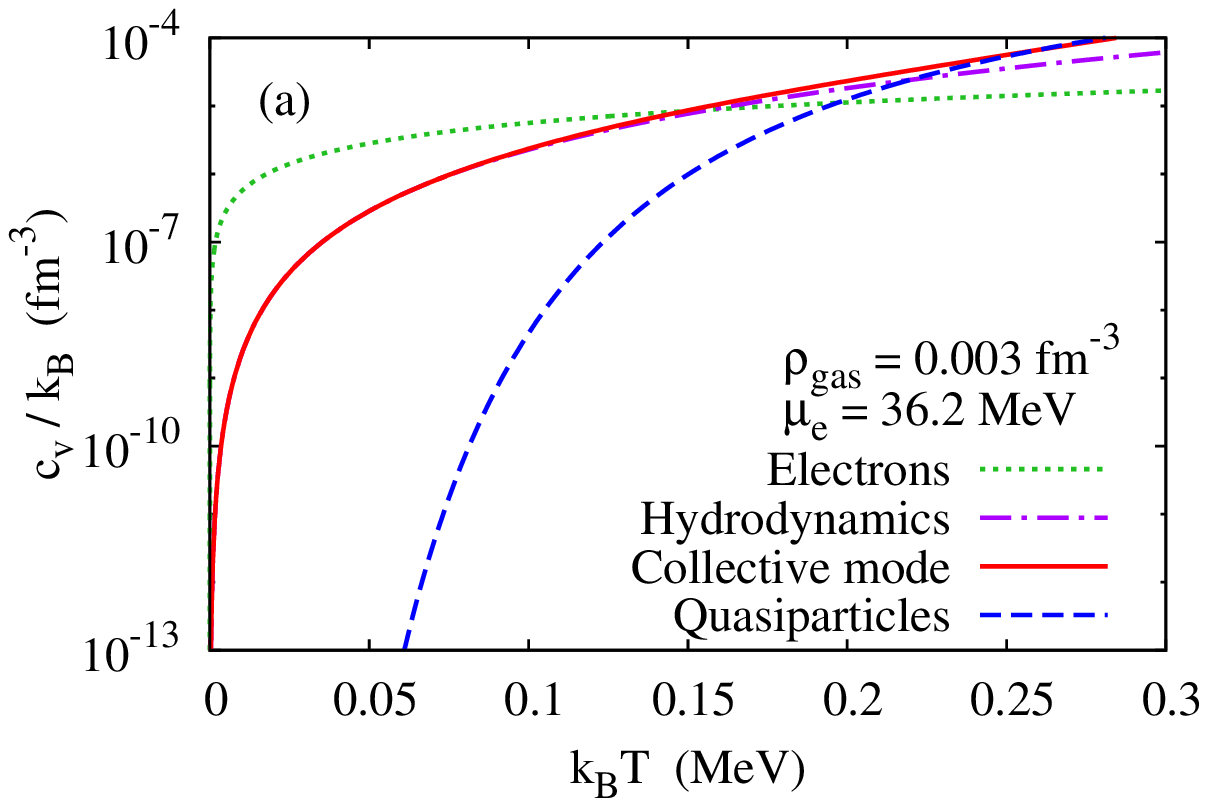}
  \includegraphics[width=8cm]{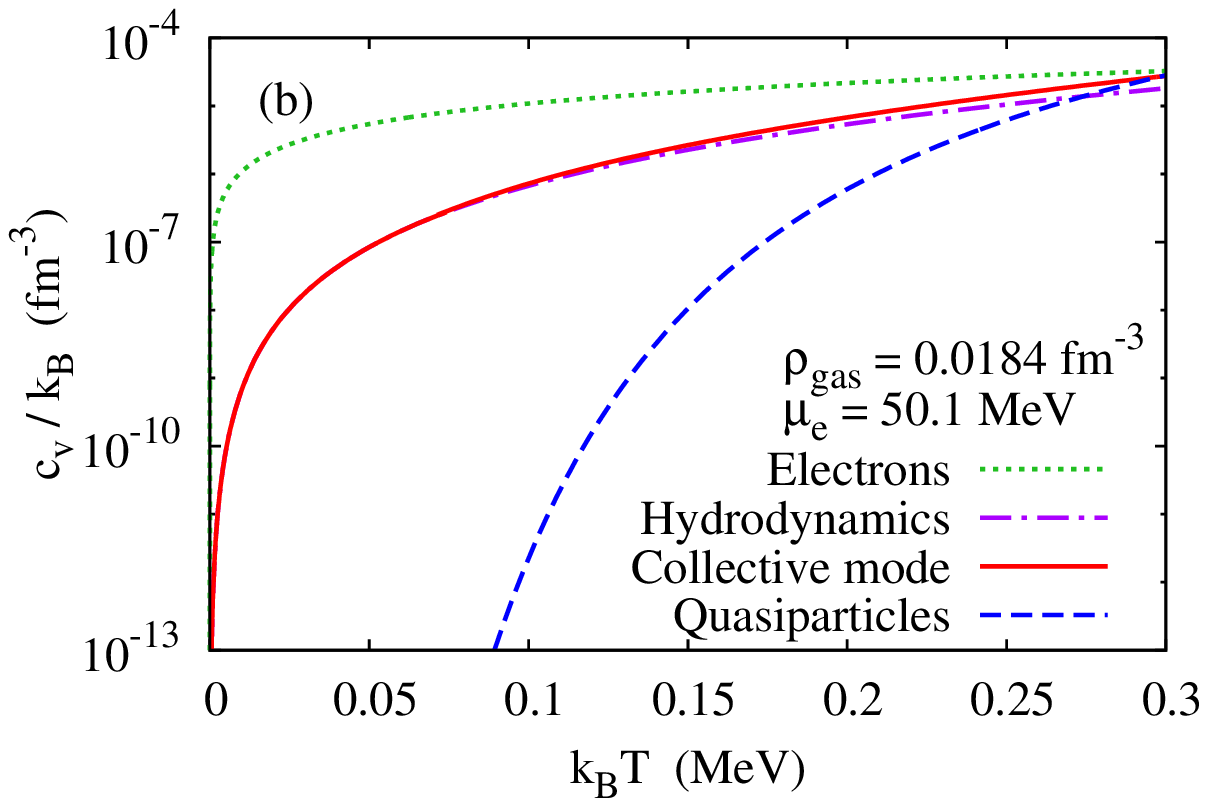}
  \caption{(Color online) Heat capacity of a neutron gas with density
    $\rho = 0.003$ (a) and $0.0184$~fm$^{-3}$ (b),
    corresponding to total baryon densities in the neutron-star crust
    of $\rho_B \approx 0.00373$ and $0.0204$~fm$^{-3}$, respectively:
    neutron quasiparticle contribution (dashes), contribution of the
    collective mode calculated within QRPA (solid lines) and within
    the hydrodynamic approximation (dashed-dotted lines). For
    comparison, we also display the electron contribution (dotted
    lines) under the assumption of $\mu_e = 36.2$ (a) and
    $50.1$~MeV~(b), corresponding to electron densities
    $\rho_e = 2.1 \cdot 10^{-4}$ and $5.5\cdot 10^{-4}$~fm$^{-3}$.
      \label{fig:cvCompare}}
   \end{figure}

At $T \ll \Delta_{k_F}$, we can neglect the temperature dependence of
the collective mode itself, i.e., we can calculate its contribution to
the specific heat by using its dispersion relation $\omega_{\vq}$
obtained at $T=0$:
\begin{equation}
   c_{v,\text{coll}} = \frac{1}{T^2}\sum_\vq \,
     \frac{\omega_\vq^2 e^{\omega_{\vq}/T}}{(e^{\omega_\vq/T} - 1)^2} \,.
\label{eq:cvQRPA}
\end{equation}
At low temperatures, this reduces to
\begin{equation}
  c_{v,\text{coll}} = \frac{2\pi^2T^3}{15u^3}\,, 
  \label{eq:cvPhonon}
\end{equation}
where $u$ is the sound velocity of the collective mode. The $T^3$
behavior is analogous to the specific heat of phonons in a solid
\cite{Debye,Ashcroft}. So, we see that at low temperatures the
contribution of the neutron gas to the specific heat is reduced as
compared to the specific heat of unpaired neutrons, which would be
linear in $T$. But the reduction is not as drastic as the exponential
suppression of $c_{v,\text{qp}}$. This is illustrated in
\Fig{fig:cvCompare},
where the specific heats of the quasiparticles, \Eq{eq:cvBCS} (dashed
lines), and of the collective mode, \Eq{eq:cvQRPA} (solid lines), are
displayed as functions of temperature. As densities of the neutron gas
we take $\rho = 0.003$ (upper panel) and $0.0184$~fm$^{-3}$ (lower
panel), which appear in the neutron-star crust at total baryon
densities of $\rho_B \approx 0.00373$ and $0.0204$~fm$^{-3}$,
respectively \cite{Negele}.

In addition to the QRPA results, we also show approximate results for
the contribution of the collective mode obtained with the hydrodynamic
sound velocity and \Eq{eq:cvPhonon} (dashed-dotted lines). At low
temperatures, \Eq{eq:cvPhonon} is in perfect agreement with the QRPA
result. This is a reassuring result since in many studies
\cite{DiGallo,Aguilera2009,Chamel2011,Cirigliano2011,
  Chamel2013,Kobyakov2013} the contribution of the collective mode was
calculated assuming the validity of the hydrodynamic approximation
(long-wavelength limit). At higher temperatures, where the QRPA result
starts to deviate considerably from \Eq{eq:cvPhonon}, also our
approximation to neglect temperature effects in the QRPA itself
becomes questionable, as one can see from the increasing contribution
of thermal quasiparticles.

Let also mention that at very low densities (such as
$\rho=0.003$~fm$^{-3}$), the sound velocity is close to that of an
ideal Fermi gas, $u \approx k_F/(\sqrt{3}m)$, so that \Eq{eq:cvPhonon}
is well approximated by $c_{v,\text{coll}} \approx 2
  \sqrt{3}m^3 T^3/(15 \rho)$. While the discrepancy between this
simple formula and \Eq{eq:cvPhonon} is less than 10\,\% in the case of
$\rho=0.003$~fm$^{-3}$, it is a factor of 3 in the case of $\rho =
0.0184$~fm$^{-3}$ where the sound velocity is considerably reduced by
the attractive neutron-neutron interaction.

To assess the importance of the contribution of the collective mode to
the specific heat of the inner crust, we show in \Fig{fig:cvCompare}
also the electron contribution (dotted lines), which is linear in
temperature,
\begin{equation}
c_{v,e} = \frac{\mu_e^2 T}{3} \, .
  \label{eq:cvel}
\end{equation}
The values of the electron chemical potentials $\mu_e = 36.2$ and
$50.1$~MeV used in the upper and lower panel of \Fig{fig:cvCompare},
respectively, were obtained from the neutron and proton chemical
potentials given in \Ref{Negele} and the relation $\mu_e =
\mu_n-\mu_p$ of $\beta$-equilibrium. One sees that, at not too low
temperatures, the contribution of the collective mode is comparable to
that of the electrons. In the case $\rho=0.003$~fm$^{-3}$, the
contribution of the collective-mode even exceeds that of the electrons
at $T \gtrsim 150$~keV.
\section{Conclusion}
\label{sec:Conclusion}
In this work we used the QRPA to study collective excitations in a
uniform superfluid neutron gas. We focused on low densities such as
they are predicted in the inner crust of neutron stars. At these
densities, the neutron pairing in the $s$ wave is relatively
strong. For the interaction, we used a Skyrme force in the ph channel
and a separable interaction with a Gaussian form factor in the pp
channel. We derived the QRPA density response by taking the
small-amplitude limit of the TDHFB equations.

Since the HFB ground state breaks the global $U(1)$ symmetry, a
Goldstone mode, corresponding to phase oscillations of the superfluid
gap, must exist. This Bogoliubov-Anderson sound is actually a simple
density wave, in other channels (e.g., spin modes) there are no
ungapped modes. Since we treat the ph and pp residual interactions
consistently with the HFB ground state, our QRPA density response
automatically exhibits the Bogoliubov-Anderson sound with a linear
dispersion relation $\omega = u q $ at low momentum $q$. The speed of
sound $u$ coincides with the hydrodynamic one. However, as $\omega$
approaches the pair-breaking threshold at $\sim 2\Delta_{k_F}$,
substantial deviations from the linear dispersion relation are found:
instead of crossing the threshold near $q=2\Delta_{k_F}/u$, the
dispersion relation of the collective mode bends, slowly approaches
the threshold, and closely follows it, before it finally crosses it at
a much higher $q$ and enters into the two-quasiparticle continuum.

We also checked the quality of the Landau approximation to the
residual interaction. We found that at low momenta ($q\lesssim
1$~fm$^{-1}$) the Landau approximation is sufficient to describe the
collective mode. In this range of momenta, also the continuum of the
response function is well described if one includes in addition to the
$l=0$ Landau parameter $F_0$ also the $l=1$ parameter $F_1$, as
required to satisfy Galilean invariance in the case of an effective
mass $m^* \neq m$. At higher momenta ($q\gtrsim 1$~fm$^{-1}$), the
QRPA response function is not well reproduced by the Landau
approximation. In this case, the inclusion of the parameter $F_1$ in
addition to $F_0$ does not significantly improve the result. However,
we note that in the case of a Skyrme interaction, the computation of
the full QRPA response is almost as simple as the calculation within
the Landau approximation, so that there is no good reason not to do
the full calculation.

The existence of an ungapped collective mode has a strong effect on
the heat capacity of the neutron gas. While quasiparticle excitations
are exponentially suppressed at low temperature $T\ll\Delta_{k_F}$
because of the gap, the collective mode can be excited at arbitrarily
low temperatures and leads to a specific heat which is proportional to
$T^3$ at low $T$, inceasing the neutron-gas contribution to the
specific heat by several orders of magnitude in the temperature range
relevant for neutron stars. Depending on density and temperature, the
contribution of the collective mode to the specific heat of the inner
neutron-star crust can be comparable to or even larger than that of
the electrons.

As we have seen, in a uniform gas the QRPA response at low energies is
well reproduced by simple hydrodynamics. However, in reality the
neutron gas in the inner crust is not uniform, but it contains
clusters having a higher density and consisting of neutrons and
protons. These clusters form a Coulomb crystal. The clusters can also
take the shape of cylinders or plates, in this case one speaks of
``pasta phases''. The coupling between the collective mode of the
neutron gas and the lattice phonons of the clusters is very important
\cite{Cirigliano2011,Chamel2013}. As long as the coherence length of
the Cooper pairs is less than the size of these structures, the
hydrodynamic approach should remain a reasonable approximation. Work
in this direction has been done in Ref.~\cite{DiGallo} for the
so-called ``lasagne'' phase and we plan to extend it to the other
geometries (crystal, ``spaghetti'' phase). For an extension of the
present study to the response of uniform matter with higher density,
as it exists in the neutron star core, one has to include also the
proton component and treat neutron pairing in the $p$ wave.

For a complete description of cooling of neutron stars
\cite{PageReddy2013}, the collective modes do not only play a role in
the specific heat, but also in the heat conductivity. A discussion of
these aspects, based on the long-wavelength approximation for the
collective modes \cite{Cirigliano2011}, can be found in
Ref.~\cite{PageReddy2012}. Again, the coupling between the collective
mode of the superfluid and the lattice phonons seems to be very
important. Therefore, a unified description of the Bogoliubov-Anderson
mode and the lattice phonons from a more microscopic perspective would
be desirable.
\acknowledgments
This work has been funded by the P2IO LabEx (ANR-10-LABX-0038) in the
framework ``Investissements d'Avenir'' (ANR-11-IDEX-0003-01) managed
by the French National Research Agency (ANR).
\appendix
\section{Skyrme parameters \label{apx:skyrmeParameters}}
In spin-unpolarized pure neutron matter, the general Skyrme functional
\cite{Brink,Engel,ChabanatSLy4} takes the particularly simple form
given in \Eq{eq:skyrmeedf}. The parameters $s_i$ are related to the
more common parameters $t_i$ and $x_i$ of \Ref{ChabanatSLy4} by
\begin{subequations}
  \begin{align}
    s_0 &= t_0 (1-x_0) \label{eq:apx:s0} \, ,\\
    s_1 &= t_1 (1-x_1) \label{eq:apx:s1} \, ,\\
    s_2 &= t_2 (1+x_2) \label{eq:apx:s2} \, ,\\
    s_3 &= t_3 (1-x_3) \label{eq:apx:s3} \, .
  \end{align}
\end{subequations}
For the numerical values of the parameters $t_i$, $x_i$, and $\alpha$,
we use the SLy4 parametrization of \Ref{ChabanatSLy4}. For
completeness, the parameters $s_i$ and $\alpha$ are listed in
Table~\ref{tab:sly4param}.
\begin{table}
\caption{Parameters of the Sly4 interaction for the case of pure
  neutron matter.
   \label{tab:sly4param}}
   \begin{ruledtabular}
   \begin{tabular}{lr}
      $s_0$ (MeV fm$^{3}$)       & -413.16\\
      $s_1$ (MeV fm$^{5}$)       &  654.29\\
      $s_2$ (MeV fm$^{5}$)       & 0\\
      $s_3$ (MeV fm$^{3+3\alpha}$) & -4877.06\\
      $\alpha$         & $1/6$
   \end{tabular}
   \end{ruledtabular}
\end{table}
Decomposing the ph interaction matrix element \Eq{eq:vphdefinition}
according to \Eq{eq:vphexpression}, one obtains:
\begin{subequations}
  \begin{gather}
     W_1(q) = s_0 + \frac{(\alpha+2)(\alpha+1)}{12} s_3 \rho^\alpha 
       + \frac{s_1-3 s_2}{4} q^2 \label{eq:w1} \, , \\
     W_2 = \frac{s_1+3 s_2}{4} \label{eq:w2} \,.
   \end{gather}
\end{subequations}

\section{Matrix of response function \label{apx:RFMatrix}}
Below we give the explicit expressions for the 16 free quasiparticle
response functions that form the matrix $\Pi^{(0)}_{\vq}$ in
\Eq{eq:respfuncQRPA}.

The $\rho^+$ response:
\begin{subequations}
  \begin{align}
    \Pi^{\rho^+,h^+}_{\vk_+,\vk_-} &= \frac{E_{\vk_+}E_{\vk_-} - \xi_{\vk_+} \xi_{\vk_-} 
      + \Delta_{\vk_+} \Delta_{\vk_-}}{4 E_{\vk_+}E_{\vk_-}} G^-_{\vk,\vq}(\omega)\,, \\
    \Pi^{\rho^+,h^-}_{\vk_+,\vk_-} &= - \frac{E_{\vk_+}\xi_{\vk_-} - \xi_{\vk_+}E_{\vk_-}}
      {4 E_{\vk_+}E_{\vk_-}} G^+_{\vk,\vq}(\omega)\,, \\
    \Pi^{\rho^+,\Delta^+}_{\vk_+,\vk_-} &= - \frac{\xi_{\vk_+} \Delta_{\vk_-}
      + \Delta_{\vk_+} \xi_{\vk_-} }{4 E_{\vk+}E_{\vk-}} G^{-}_{\vk,\vq}(\omega)\,, \\
    \Pi^{\rho^+,\Delta^-}_{\vk_+,\vk_-} &= - \frac{E_{\vk_+} \Delta_{\vk_-} 
      + \Delta_{\vk_+} E_{\vk-}}{4 E_{\vk+}E_{\vk-}} G^+_{\vk,\vq}(\omega)\,.
  \end{align}
The $\rho^-$ response:
  \begin{align}
    \Pi^{\rho^-,h^+}_{\vk_+,\vk_-} &= - \frac{E_{\vk_+} \xi_{\vk_-} - \xi_{\vk_+} E_{\vk_-}}
      {4 E_{\vk_+}E_{\vk_-}} G^+_{\vk,\vq}(\omega)\,, \\
    \Pi^{\rho^-,h^-}_{\vk_+,\vk_-} &= \frac{E_{\vk_+} E_{\vk_-} - \xi_{\vk_+} \xi_{\vk_-}
      - \Delta_{\vk_+} \Delta_{\vk_-}}{4 E_{\vk_+}E_{\vk_-}} G^-_{\vk,\vq}(\omega)\,, \\
    \Pi^{\rho^-,\Delta^+}_{\vk_+,\vk_-} &= - \frac{E_{\vk_+} \Delta_{\vk_-}
      - \Delta_{\vk_+} E_{\vk_-}}{4 E_{\vk_+}E_{\vk_-}} G^+_{\vk,\vq}(\omega)\,, \\
    \Pi^{\rho^-,\Delta^-}_{\vk_+,\vk_-} &= - \frac{\xi_{\vk_+} \Delta_{\vk_-} 
      - \Delta_{\vk_+} \xi_{\vk_-}}{4 E_{\vk_+}E_{\vk_-}} G^-_{\vk,\vq}(\omega)\,.
  \end{align}
The $\kappa^+$ response:
  \begin{align}
    \Pi^{\kappa^+,h^+}_{\vk_+,\vk_-} &= \frac{\xi_{\vk_+} \Delta_{\vk_-} 
      + \Delta_{\vk_+} \xi_{\vk_-}}{4 E_{\vk_+}E_{\vk_-}} G^-_{\vk,\vq}(\omega)\,, \\
    \Pi^{\kappa^+,h^-}_{\vk_+,\vk_-} &= \frac{E_{\vk_+} \Delta_{\vk_-} 
      - \Delta_{\vk_+} E_{\vk_-}}{4 E_{\vk_+}E_{\vk_-}} G^+_{\vk,\vq}(\omega)\,, \\
    \Pi^{\kappa^+,\Delta^+}_{\vk_+,\vk_-} &= - \frac{E_{\vk_+} E_{\vk_-} 
      + \xi_{\vk_+} \xi_{\vk_-} - \Delta_{\vk_+} \Delta_{\vk_-}}{4 E_{\vk+}E_{\vk-}}
      G^-_{\vk,\vq}(\omega)\,, \\
    \Pi^{\kappa^+,\Delta^-}_{\vk_+,\vk_-} &= - \frac{E_{\vk_+} \xi_{\vk_-} 
      + \xi_{\vk_+} E_{\vk_-}}{4 E_{\vk_+}E_{\vk_-}} G^+_{\vk,\vq}(\omega)\,.
  \end{align}
The $\kappa^-$ response:
  \begin{align}
    \Pi^{\kappa^-,h^+}_{\vk_+,\vk_-} &= \frac{E_{\vk_+} \Delta_{\vk_-} 
      + \Delta_{\vk_+} E_{\vk_-}}{4 E_{\vk_+}E_{\vk_-}} G^+_{\vk,\vq}(\omega)\,, \\
    \Pi^{\kappa^-,h^-}_{\vk_+,\vk_-} &= \frac{\xi_{\vk_+} \Delta_{\vk_-} 
      - \Delta_{\vk_+} \xi_{\vk_-}}{4 E_{\vk_+}E_{\vk_-}} G^-_{\vk,\vq}(\omega)\,, \\
    \Pi^{\kappa^-,\Delta^+}_{\vk_+,\vk_-} &= - \frac{E_{\vk_+} \xi_{\vk_-} 
      + \xi_{\vk_+} E_{\vk_-}}{4 E_{\vk_+}E_{\vk_-}} G^+_{\vk,\vq}(\omega)\,, \\
    \Pi^{\kappa^-,\Delta^-}_{\vk_+,\vk_-} &= - \frac{E_{\vk_+} E_{\vk_-} 
      + \xi_{\vk_+} \xi_{\vk_-} + \Delta_{\vk_+} \Delta_{\vk_-}}{4 E_{\vk_+}E_{\vk_-}} 
      G^-_{\vk,\vq}(\omega)\,.
  \end{align}
  \label{eq:RFexplicit}
\end{subequations}

In the above expressions we have used the abbreviation
\begin{equation}
G^\pm_{\vk,\vq}(\omega) = \frac{1}{\omega - \Omega_{\vk,\vq} + i \eta} 
  \pm \frac{1}{\omega + \Omega_{\vk,\vq} + i \eta} \, ,
  \label{eq:apx:gpm}
\end{equation}
where $\Omega_{\vk,\vq} = E_{\vk_+} + E_{\vk_-}$.

The matrix $\sumk{\Pi^{(0)}_{\vek{q}}V}$ used in \Eq{eq:qrpaRF} is defined as
\begin{widetext}
   \begin{multline}
      \sumk{\Pi^{(0)}_{\vek{q}}V} = \\
       W_1(q)
      \begin{pmatrix}
         \sumk{\Pi^{\rho^+,h^+}_{\vk_+,\vk_-}} & 0& 0& 0& 0 \\
         \sumk{k^2 \Pi^{\rho^+,h^+}_{\vk_+,\vk_-}} & 0& 0& 0& 0 \\
         \sumk{k z \Pi^{\rho^-h^+}_{\vk_+\vk_-}} & 0& 0& 0& 0 \\
         \sumk{F(k) \Pi^{\kappa^+,h^+}_{\vk_+,\vk_-}} & 0& 0& 0& 0 \\
         \sumk{F(k) \Pi^{\kappa^-,h^+}_{\vk_+,\vk_-}} & 0& 0& 0& 0 \\
      \end{pmatrix}
      + W_2
      \begin{pmatrix}
         \sumk{k^2 \Pi^{\rho^+,h^+}_{\vk_+,\vk_-}} & 
         \sumk{\Pi^{\rho^+,h^+}_{\vk_+,\vk_-}} &
         - 2 \sumk{k z \Pi^{\rho^+,h^-}_{\vk_+,\vk_-}} &
         0 &0 \\
         \sumk{k^4 \Pi^{\rho^+,h^+}_{\vk_+,\vk_-}} & 
         \sumk{k^2 \Pi^{\rho^+,h^+}_{\vk_+,\vk_-}} & 
         - 2 \sumk{k^3 z \Pi^{\rho^+,h^-}_{\vk_+,\vk_-}} &
         0 &0 \\
         \sumk{k^3 z \Pi^{\rho^-,h^+}_{\vk_+,\vk_-}} &
         \sumk{k z \Pi^{\rho^-,h^+}_{\vk_+,\vk_-}} &
         - 2 \sumk{k^2 z^2 \Pi^{\rho^-,h^-}_{\vk_+,\vk_-}} &
         0 &0 \\
         \sumk{F(k) k^2 \Pi^{\kappa^+,h^+}_{\vk_+,\vk_-}} &
         \sumk{F(k) \Pi^{\kappa^+,h^+}_{\vk_+,\vk_-}} &
         - 2 \sumk{F(k) k z \Pi^{\kappa^+,h^-}_{\vk_+,\vk_-}} &
         0 &0 \\
         \sumk{F(k) k^2 \Pi^{\kappa^-,h^+}_{\vk_+,\vk_-}} &
         \sumk{F(k) \Pi^{\kappa^-,h^+}_{\vk_+,\vk_-}} &
         - 2 \sumk{F(k) k z \Pi^{\kappa^-,h^-}_{\vk_+,\vk_-}} &
         0 &0 
      \end{pmatrix} \\
      + g
      \begin{pmatrix}
         0&0&0&
         \sumk{F(k) \Pi^{\rho^+,\Delta^+}_{\vk_+,\vk_-}} &
         \sumk{F(k) \Pi^{\rho^+,\Delta^-}_{\vk_+,\vk_-}} \\
         0&0&0&
         \sumk{F(k) k^2 \Pi^{\rho^+,\Delta^+}_{\vk_+,\vk_-}} &
         \sumk{F(k) k^2 \Pi^{\rho^+,\Delta^-}_{\vk_+,\vk_-}} \\
         0&0&0&
         \sumk{F^2(k) \Pi^{\kappa^+,\Delta^+}_{\vk_+,\vk_-}} &
         \sumk{F^2(k) \Pi^{\kappa^+,\Delta^-}_{\vk_+,\vk_-}} \\
         0&0&0&
         \sumk{F^2(k) \Pi^{\kappa^+,\Delta^+}_{\vk_+,\vk_-}} &
         \sumk{F^2(k) \Pi^{\kappa^+,\Delta^-}_{\vk_+,\vk_-}} \\
         0&0&0&
         \sumk{F^2(k) \Pi^{\kappa^-,\Delta^+}_{\vk_+,\vk_-}} &
         \sumk{F^2(k) \Pi^{\kappa^-,\Delta^-}_{\vk_+,\vk_-}} 
      \end{pmatrix}
      \, ,
  \label{eq:QRPAmatrix}
  \end{multline}
\end{widetext}
with $z = \cos \sphericalangle(\vk,\vq)$. 
\section{Numerical computation \label{apx:numComp}}
In Appendix~\ref{apx:RFMatrix} we gave the equations needed to
determine the QRPA response function. In practice, the summations over
$\vk$ are integrals. In our numerical calculations we start by
evaluating the imaginary parts of the matrix
$\sumk{\Pi^{(0)}_\vq V}$. According to \Eqs{eq:RFexplicit} and
(\ref{eq:QRPAmatrix}), each element of this matrix can be written in
the form
\begin{equation}
   \sumk{\Pi^{(0)}_{\vq}(\omega) V}_{\alpha\beta} = \int \frac{d^3k}{(2\pi)^3}
     f_{\alpha\beta}(k,q,z) G^\pm_{\vk,\vq}(\omega) \, .
   \label{eq:RFGeneralForm}
\end{equation}
Then the the imaginary part is given by :
\begin{equation}
   \Imag \sumk{\Pi^{(0)}_{\vq}(\omega)V} =
   \frac{1}{2 \pi^2} \int_0^{z_\maxi} dz \sum_i \frac{k_i^2 f(k_i,q,z)}
     {\left| \frac{\partial \Omega_{\vk,\vq}}{\partial k}\right|_{k_i}}
     \, ,
   \label{eq:apx:imaginarySum}
\end{equation}
where $\{k_i\}$ is the set of solutions of the equation
$\Omega_{\vk,\vq}=\omega$ for a given angle $z$, and $z_{\maxi}$ is
either 1 or the angle beyond which the equation
$\Omega_{\vk,\vq}=\omega$ does not have a solution any more. After the
calculation of the imaginary part, we compute the real part with the
help of a dispersion relation,
\begin{multline}
   \Real \sumk{\Pi^{(0)}_{\vq}(\omega)V} =
   - \frac{1}{\pi} \int_0^\infty d\omega' \Imag \sumk{\Pi^{(0)}_{\vq}(\omega')V}
   \\ \times
   \left( \frac{1}{\omega-\omega'} \pm \frac{1}{\omega+\omega'} \right) \, ,
   \label{eq:apx:realSum}
\end{multline}
where the sign $\pm$ is chosen according to the sign in $G^\pm$ in
\Eq{eq:RFGeneralForm}.
\nocite*
\bibliographystyle{apsrev4-1}
\bibliography{qrpa_article}

\end{document}